# Relational Quantum Mechanics, Quantum Relativism, and the Iteration of Relativity

Timotheus Riedel

**ABSTRACT**. The idea that the dynamical properties of quantum systems are invariably relative to other systems has recently regained currency. Using Relational Quantum Mechanics (RQM) for a case study, this paper calls attention to a question that has been underappreciated in the debate about quantum relativism: the question of whether relativity iterates. Are there absolute facts about the properties one system possesses relative to a specified reference, or is this again a relative matter, and so on? It is argued that RQM (in its best-known form) is committed to what I call the Unrestricted Iteration Principle (UIP), and thus to an infinite regress of relativisations. This principle plays a crucial role in ensuring the communicability and coherence of interaction outcomes across observers. It is, however, shown to be incompatible with the widespread, conservative reading of RQM in terms of relations, instead necessitating the adoption of the more unorthodox notion of perspectival facts. I conclude with some reflections on the current state of play in perspectivist versions of RQM and quantum relativism more generally, underscoring both the need for further conceptual development and the importance of the iteration principle for an accurate cost-benefit analysis of such interpretations.

# 1  Introduction

Recent years have seen a fresh trend towards relativism in the foundations of quantum theory. That the dynamical states of quantum systems are in some interestingly novel sense *relative* is, of course, a suspicion much older than that. In very different guises, it can be rediscovered in the writings of towering figures like Bohr, Hermann, and Everett. Its new upswing owes not least to a series of thought experiments (so-called "Bell-Wigner mashups" (Leifer 2018)) that purportedly give rise to no-go theorems for observer-independent measurement outcomes in single-world interpretations (e.g., Brukner 2017; Frauchiger & Renner 2018; Bong et. al. 2020).

Relational Quantum Mechanics (RQM) is an interpretive programme, initiated by physicist Carlo Rovelli about 25 years ago, that has been at the forefront of the relativist renaissance. It was founded on the hypothesis that state-dependent properties of physical objects are invariably relative to reference systems. Lately, RQM has been much-discussed, applied to different conundrums in the quantum domain, entertained as a conceptual stepping stone on the road to quantum gravity, but also vehemently criticised.

The purpose of this paper is to draw attention to a principle that plays a role both essential and precarious in RQM but whose impact has so far been underappreciated. I call it the *Unrestricted Iteration Principle (UIP)*, or the *iteration principle* for short. Once it has been claimed that properties are relative, the question arises whether there are *absolute* facts about what the properties are relative to some particular reference. According to the iteration principle, there are not: instead, this second-order matter is again relative to reference systems; and likewise for third-order matters, fourth-order matters, and so on. Whether or not to embrace this iteration of relativity is, I believe, a consequential question for *any* relativist approach to quantum theory. To no small extent, both the considerations that speak in its favour and the consequences of its acceptance play out similarly in a broad class of such approaches. In this sense, my present interest in RQM is partly as a case study for what I suspect to be a more general dialectic.

In the first half of the paper (section 2), I introduce RQM and explain how the iteration principle emerges naturally (albeit not inevitably) from its core commitments. The principle is shown to play an integral role in Rovelli's story, underwriting the epistemic accessibility of interaction events to outside observers whilst bypassing the threat of cross-perspective incoherences. I also place in context a substantially modified version of RQM recently entertained by Adlam & Rovelli (2023), which does *not* rely on UIP but significantly cuts down on the relativist flavour of the interpretation.

The second part investigates into the wider ramifications of the iteration principle. In section 3.1, I distinguish two forms of relativism about a domain: relationalism and realism about perspectival facts ('perspectivism'). Which of them the relativity of properties is to be modelled on is a question the literature on RQM has been ambiguous about. Many commentators presuppose the less extravagant relationalist paradigm. But a simple argument presented in section 3.2 establishes that the combination of relationalism and UIP is unpalatable, for it implies that nothing (in the domain at issue) is the case at all. It follows that one is either forced to hedge in the iteration principle, and in a way that does not betray its original motivation, or stuck with the more outlandish notion of perspectival facts.

I conclude with some reflections on the current state of play in perspectivist readings of RQM in section 3.3, finding that they have not yet been developed in sufficient conceptual detail. For one thing, engagement with the extant metaphysical literature and the different possible *types* of perspectivism it has identified (cf. Fine 2005) seems advisable. For another, disagreement appears to persist about the fate of the iteration principle in perspectivist RQM. If the principle is rejected, the task again is to find a workable restriction. If it is embraced, perspectivist RQM implies an infinite regress of perspective-dependencies, which some have found undesirable, untenable, or even incoherent. I do not put these considerations forward as fatal objections, but to highlight a number of pending questions for advocates of perspectivist RQM, or indeed perspectivist quantum theory more generally.

## 2  Radical Rovellian Relativity

### 2.1  Wigner's Friend, Relatively Easy

Similar to its rival QBism, Relational Quantum Mechanics has not yet assumed a definite shape, and is in this sense aptly characterised as an ongoing programme. Here, we begin with a reconstruction of the ideas distinctive of what one might call 'classical' RQM, prefigured in Rovelli's seminal (1996) and further matured by the time of Rovelli (2022[a]) and Di Biagio & Rovelli (2022). This version is the main object of study for the present paper. At the appropriate moment, I will also consider an alternative variety more recently entertained by Adlam & Rovelli (2023).

To a good approximation, classical RQM is the attempt of interpreting quantum theory through the hypothesis that physical systems only ever possess state-dependent properties during interactions with and relative to one another. The original motivation for this working assumption stemmed from the famous 'Wigner's friend' thought experiments (cf. Rovelli 1996, 1642-1644). In a nutshell, these illustrate that two competent users of 'textbook quantum theory' initially in agreement about the state of some system may end up assigning different, apparently incompatible states to it. For vividness, suppose that in a perfectly isolated laboratory, Friend performs an (ideal, non-disturbing) measurement of a two-valued variable q with possible values {*up*, *down*} on a system S whose initial quantum state has form $\alpha |S_{up}\rangle + \beta |S_{down}\rangle$. According to the textbooks, Friend witnesses one of the possible outcomes, and S's quantum state collapses onto the appropriate eigenstate. Famously, though, a second observer Wigner who does not interact with either Friend or S would model the goings-on quite differently; namely, as an uninterrupted unitary evolution of their combined quantum state of the type

$$(\alpha |S_{up}\rangle + \beta |S_{down}\rangle) \otimes |F_{ready}\rangle \rightarrow \alpha (|S_{up}\rangle \otimes |F_{up}\rangle) + \beta (|S_{down}\rangle \otimes |F_{down}\rangle),$$

where $F_{ready}$ is Friend's initial state, and $F_{up/down}$ describe her as having observed the outcome *up* or *down*, respectively. At time t when Friend has completed her measurement, she ascribes either the *up*- or the *down*-property to S; yet Wigner's description rather seems to suggest that q does not have any one particular value at all. Seen from this angle, it looks almost as if quantum theory, when interpreted in a broadly realistic spirit and employed by multiple observers, threatens to be inconsistent.

One is naturally tempted to put the mismatch down to a lack of knowledge on Wigner's part, and consequently regard Friend's account as privileged. After all, Friend holds information (about her measurement outcome) currently unavailable to Wigner. On this view, Wigner's state ascription is either in some sense *incorrect* or at least *incomplete*. However, as is well-known, this ignorance-based reading of the case is not entirely straightforward. For instance, Wigner should in principle be in a position to witness interference effects. In more general terms, the superposition he ascribes to the lab does not give rise to the same predictions as a proper mixture (cf. d'Espagnat 1966).

In light of this, Rovelli argued that Friend's and Wigner's accounts should both be treated as correct and complete. Consequently, he proposed to accept what seems to be suggested by their divergent quantum state assignments – namely, that at time t, q does have a determinate value relative to Friend but not relative to Wigner. He concluded that quantum states as well as state-dependent properties of physical systems ought to be regarded as relative to other systems (1996, 1650). This marked the birth of Relational Quantum Mechanics. More recently, the relativity of dynamical properties has, in subtly different ways[1], also been advocated by Neo-Copenhagenists (Auffèves & Grangier 2016; Brukner 2017), Healey (2022), Dieks (2022), and some QBists (cf. Pienaar 2021[b]), amongst others. As indicated above, this development has in part been nurtured by a number of technical results that purportedly cast doubt on the existence of observer-independent facts about measurement outcomes in single-world interpretations of quantum mechanics (e.g., Brukner 2017; Frauchiger & Renner 2018; Bong et. al. 2020).

Rovelli has always underlined that agents, consciousness, or 'measurements' play no special role on his view (ibid., 1641). Rather, systems acquire properties relative to any other system they physically interact with, be it a qubit, a milkshake, or a human: a variable of some A spontaneously assumes a determinate (relative) value when it gets "appropriately entangled" (Di Biagio & Rovelli 2022, 14) with a variable of some B. Such a property manifestation is also called a (relative) 'event' (cf. Rovelli 2022[a], 1055). Over time, the idea emerged that systems only ever possess determinate properties *at*, but not in between, interactions. Thus, a physical process is to be thought of as "a very fine-grained but discrete swarming" (Rovelli 2018, 9) of flash-like events, instead of a continuous evolution. This, Rovelli argues, is a plausible conclusion to draw both from the Kochen-Specker theorem (ibid.) and the conceptual thinking that as a matter of historical fact led Heisenberg to the development of matrix mechanics (Rovelli 2022[a], 1057).

---

1 Auffèves & Grangier take properties to be relative to the *context* in which a system is embedded; Healey construes them as relative to *decoherence environments*. For the QBists and Brukner, only *agents* or *observers*, in a substantial sense of that term, qualify as reference systems. Dieks assumes a more pronouncedly realist attitude towards quantum states than the others.

While variables and the values (or, more accurately, determinables and the determinates) they acquire in interactions loom large in the ontology of classical RQM, the significance of the quantum state has remained volatile and controversial. What is clear is that it is not taken to represent a physical *object*. According to recent expositions, it is interpreted "epistemically" (Rovelli 2021[a], 2), as a mere bookkeeping device that takes in known events between two systems to output probabilistic predictions for other interactions between them (2022[a], 1061). This deflated reading plays an essential role in extant arguments to the effect that RQM preserves locality (Smerlak & Rovelli 2007; Martin-Dussaud, Rovelli & Zalamea 2019), and in the rebuttal (Di Biagio & Rovelli 2022, 7-9) of certain recent objections due to Pienaar (2021[a]). Nonetheless, how we are to think of the probabilities encoded in the state vector (as ontic, epistemic and objective, or even entirely subjective) remains ambiguous.[2] This would matter greatly if we wanted to assess Rovelli's suggestion that quantum states in RQM are not descriptive of physical reality (cf. Stacey 2021, 3-5).

## 2.2 Iterating Relativity

From the preceding, it may not yet be obvious why RQM should be considered particularly radical. History has got us used to the unexpected relativisation of observables – spatial distances, time intervals, you name it. Indeed, it has sometimes been claimed that RQM is "not so unconventional" (Vidotto 2002, 164), but just another instance of a broad trend towards relationality in modern physics (ibid.; cf. Martin-Dussaud 2021). However, as recently stressed by Pienaar (2021[a], 7-10) and acknowledged by Di Biagio & Rovelli (2022, 6), the analogy only goes so far. While the aforementioned magnitudes are indeed relative to reference frames in relativity theory, there always is an *absolute* fact of the matter about their value as is relative to any specified frame. If we want, we can pass from unary ('has length $x$', say) to binary predicates ('has length $x$ relative to frame $f$') to recover statements of absolute validity. It emerges from Rovelli's writings, though, that this is not the case in classical RQM: there is *no* absolute fact about the properties a system possesses relative to a particular other. Instead, relativity *iterates*. Originally, this assumption arose from the idea that a fact about a property which one system possesses relative to another must itself be a fact about the dynamical state of their composite:

> "If the statement '$q$ has a value relative to $O$' [...] has any comprehensible physical meaning at all, this meaning should be related to the contingent state of the *S-O* system. According to the main hypothesis here, asking about the observer-independent contingent state of the S-O system has no meaning [...] [But] we can make statements about the state of the S-O system, provided that we interpret these statements as relative to a third physical system P" (Rovelli 1996, 1653).

---

2  Smerlak & Rovelli (2007, 431) claim that quantum probabilities in RQM are "clearly to be interpreted subjectively". Elsewhere, Rovelli says that transition probabilities provide "something objective" (Smerlak et. al. 2017, 221). On an alternative proposal by Dorato (2016, 242), cited favourably in Laudisa & Rovelli (2019, 3.2), state vectors represent ontic probabilistic dispositions of systems. Pienaar (2021[b], 5f.) argues that RQM *requires* ontic probabilities. Weststeijn thinks that "RQM does not treat the quantum state as something that is purely epistemic" (2021, 5).

Correspondingly, Rovelli takes the relative properties between two systems to be rather directly represented by state-dependent variables of their composite (whose values, once again, are likewise relative as per the 'main hypothesis'). Reconsider our earlier thought experiment: At t, the quantum state Wigner assigns to the target system S and his friend F is of form

$$\alpha \, (|S_{up}\rangle \otimes |F_{up}\rangle) + \beta \, (|S_{down}\rangle \otimes |F_{down}\rangle).$$

Now, suppose Wigner desires to learn whether the observable assumed value *up* relative to Friend. According to Rovelli, Wigner can simply interact with the composite S+F to measure whether it is in state $|S_{up}\rangle|F_{up}\rangle$ (viz., check target and/or pointer variable) (cf. Rovelli 1996, 1651f.). But this interaction is *itself* the measurement of a (two-valued) state-dependent observable on S+F, whose value must by assumption be relative. After all, this observable could itself be plugged into a Wigner's friend scenario. (We did not make any assumptions about the internal structure of S when we inferred that its property is relative – it could itself well be a composite of two subsystems.) This led Rovelli to assert that the notion that one system has a property relative to another "is a physical notion that can be studied experimentally (by a third observer) in the same way as any other physical property of a system" (ibid., 1666).

Rovelli and collaborators repeatedly reaffirmed that "there is no absolute way of comparing the perspectives of two systems" (Di Biagio & Rovelli 2022, 2), such that "[i]t is meaningless to compare events relative to different systems, unless this is done relative to a (possibly third) system" (ibid., 4; cf. also Smerlak & Rovelli 2007, 441; Rovelli 2022$^a$, 1066).

It is important to understand that second-order relativity serves at least two dialectical purposes in classical RQM. Of those, we have already touched upon one: by associating facts about the properties of one system relative to another with a state-dependent variable of the composite, one hopes to render them epistemically accessible to outside observers. This has always been a principal concern for Rovelli (1996, 1651f.; 2022$^a$, 1066). Arguably, the only sensible thing Wigner could do to inquire into the interaction event between S and Friend is 'measure' state-dependent variables of other physical systems, above all S and Friend themselves. If there were an *absolute* fact about the property S assumed relative to Friend, it is unclear how Wigner could infer it from the values some variables assume *relative to him*.[3] To connect these two things, one would have to provide some additional theoretical structure RQM in its original form does not provide.[4]

But what is more, second-order relativity also is key in Rovelli's response to a natural concern about the consistency of events across perspectives. If physical objects only have relative variables, could they not assume wildly different and incompatible properties towards different observers?[5] This would insofar be problematic as (discounting errors and the like) nothing of the sort ever seems to be witnessed in real-world laboratories. Otherwise, the relativity of properties would presumably have been discovered a long time ago. If the interpretation would allow for such discrepancies, it would threaten to be empirically falsified, or else would have to systematically hide the phenomenon from view somehow. In the worst case, that could mean that observers live in mutually disconnected realities.[6]

---

3   This is why some of those who pass over the iteration of relativity (or its importance) have concluded that external observers cannot find out what happened in an interaction (cf. Healey 2022, 15; Muciño, Okon & Sudarsky 2022, 18f.). Pienaar (2021$^a$, 16-20), Adlam (2022, 10), and Dieks (2022, 12f.) also notice this dialectic.
4   The lesson strikes me as more general: If one regards state-dependent properties as relative (for instance because one is convinced by one of the no-go theorems), one is challenged to sail between the Scylla of epistemic inaccessibilities and the Charybdis of a regress of relativity (cf. also Dieks 2022, 12f.; Adlam 2022).
5   This is sometimes framed as the worry of whether two observers might see different outcomes when they successively measure the same variable. But of course, measurements need not be repeatable, anyhow. At core, the question is whether some S may manifest properties in its interactions with some S' that would be considered impossible (or astronomically unlikely) by the lights of its quantum state relative to some other S''.

Second-order relativity allows Rovelli to sidestep this set of problems. For, it implies that there *is no absolute fact* about the value of a variable relative to some particular reference. Accordingly, "the question 'Do observers *O* and *P* get the *same* answers out of a system *S*?' is a meaningless question, because it is a question about the *absolute* state of O and P" (1996, 1666; original emphasis; cf. the same argument in Smerlak & Rovelli 2007, 441; Rovelli 2021[a], 6; Di Biagio & Rovelli 2022, 2-4).[7] One may perhaps still wonder whether it is possible that *relative to some reference*, two systems make incompatible observations. But this, Rovelli argues, is ruled out by the quantum formalism itself (1996, 1652; Smerlak & Rovelli 2007, 439). To give a simple example, suppose that Wigner has *two* friends F1, F2 inside the lab, who both measure variable q. In this case, the state he would assign at t would have the form

$$\alpha \left(|S_{up}\rangle \otimes |F1_{up}\rangle \otimes |F2_{up}\rangle\right) + \beta \left(|S_{down}\rangle \otimes |F1_{down}\rangle \otimes |F2_{down}\rangle\right).$$

Since q and the two pointer variables are correctly correlated within each term of this superposition, Wigner could be certain that his friends would agree about the value of q if he were to ask them. Summing up: Where the claim that different observers may receive incompatible measurement results is intended as absolute, it is ill-posed; and where it is intended as relative, it is false.

Assume, then, that we accept the second layer of relativity. In this case, we can repeat the same line of reasoning on the next-higher level: The value of S's variable relative to Friend *relative to Wigner* corresponds to a state-dependent property of the composite S+F+W. If we knew that Wigner had interacted with S+F, we could find out whether the observable assumed value *up* relative to Friend relative to Wigner by checking whether S+F+W is in state $|S_{up}\rangle|F_{up}\rangle|W_{up}\rangle$. Supposedly, this ensures that the interaction between S+Friend and Wigner is epistemically accessible to us. We conclude that there is no absolute fact about the property which S acquired relative to Friend relative to Wigner. Again, this is only for the better: if there were, we would get to wonder whether S could have taken on an entirely different property relative to Friend relative to somebody *else*.

---

6  Not that empirically inaccessible states of affairs are forbidden *per se* – other interpretations (like de Broglie-Bohm theory) have them, too. But one would have to *explain* these epistemic barriers, and without proving those right who worry that RQM locks each system in its own perspective (cf. Pienaar 2021[a], 20f.; Adlam 2022; Healey 2022, 15; Muciño, Okon & Sudarsky 2022, 17-22).

7  Adlam (2022, 9f.) argues that this solution is problematic, because scientific confirmation *requires* that the question be well-posed (this is the main motivation for the new version of RQM developed in Adlam & Rovelli, 2023). Healey (2022, 15f.) agrees, while Dieks (2022, 12) dissents.

Of course, we now seem to have kick-started a regress (as noted by Muciño, Okon & Sudarsky (2022, 13) and anticipated by Van Fraassen (2010, 415f.)). Pending additional considerations, it is reasonable to accept the same argument on the third, fourth, or indeed any level if at all. This would seem to extend even to *infinite* sequences of *relata*: The value of some variable of $S_1$ relative to $S_2$ relative to $S_3$ ... [ad inf.] would correspond to a state-dependent property of the composite constituted by the $S_i$, which by hypothesis is relative, too. No matter the ordinality of the sequence of *relata* we amass, then, we would end up with something that is itself relative; or, turning things around, there would be no way of regaining absolutely valid statements by specifying ever more *relata*. Call this the *Unrestricted Iteration Principle*, or *UIP* for short.

Although they never addressed the question of third- or higher-order relativity explicitly, the writings of Rovelli and collaborators indicate that classical RQM was meant to include UIP. According to Rovelli's 'main hypothesis', "the concept of a description of a system independent of the observer" (1996, 1669) ought to be altogether abandoned. But insofar as the description of the properties some system possesses relative to a sequence of other systems is a description of the state of their composite, halting the iteration of relativity on any level would be tantamount to allowing some absolutely valid description of the dynamical state of some system. This is, of course, just a generalisation of Rovelli's own argument, quoted above, from the case of second-order relativity to arbitrary orders. Note further that he and collaborators insisted that *whatever* is said about the dynamical states of physical systems must be indexed to a reference, and that this principle should be understood to extend even to their states relative to each other (Smerlak & Rovelli 2007, 441; Rovelli 2021[a], 6). Consequently, no view from nowhere can be recovered in RQM (Di Biagio & Rovelli 2022, 11). All this is also in line with the take Rovelli offers on the teachings of Nāgārjuna as a possible philosophical home for RQM in his popular science book *Helgoland*.[8] Therein, Rovelli suggests that everything, including relational structures themselves, "exists only [...] in relation to something else" (2021[b], 127). Perhaps most explicitly, Adlam & Rovelli have recently declared that in classical RQM, "even facts about relations must always be relativized to an observer" (2023, 3).

For the rest of this paper, I assume that UIP forms part of the most plausible reconstruction of classical RQM. If so, it is certainly one of the most challenging and most striking features of the interpretation. Remarkably, however, the iteration of relativity has frequently been ignored or overlooked. This is true even for second-order relativity, which Rovelli and collaborators promoted comparatively explicitly. Sometimes, it is claimed that relativity does *not* iterate in RQM (Dorato 2016, 252; cf. also Van Fraassen 2010, 399).[9] But usually, the matter is simply not touched upon.[10] The point generalises: One would be hard pressed to find an explicit discussion of iterated relativity anywhere in the literature on quantum relativism. I believe that this lack of attention has had a hampering influence on the debate. Not only has it led some commentators, both in the debate about RQM proper and beyond, to miss the importance of the question for issues of cross-perspective communication and consistency (cf. Muciño, Okon & Sudarsky 2022, 17-19; Healey 2022, 15; Dieks 2022, 12). As I argue below, it has also abetted a fairly widespread yet mistaken reading of classical RQM.

Three points merit attention before closing this section.

First, there is an important related question about whether systems can show up in a sequence of *relata* more than once. Originally, Rovelli was inclined to think that they could not (1996, 1669-1672). This was based on the idea that "there is no meaning in being correlated with oneself" (ibid., 1666), and on Breuer's theorem (1995) to the effect that systems cannot measure their own dynamical states completely. However, if we do not allow for repetitions in the sequence, then any attribution of a state-dependent property would (by UIP) have to rely on an inexhaustible pool of distinct reference systems. This would, for instance, mean that I could only ever possess a dynamical property relative to a never-ending series of different systems, and only if all these infinitely many systems had appropriately interacted with me (at least indirectly) such as to correlate their own states with said property of mine. If we assume that the number of systems in Friend's (perfectly isolated) lab is finite, it actually *could not be the case* that S or Friend had any state-dependent property at t whatsoever, because there would not be enough systems it could be the case relative to.

---

8   Many thanks to an anonymous referee whom I owe this observation to.
9   Dorato later corrected this verdict (2020, 241).
10  Examples abound, but to name a few: Candiotto 2017; Laudisa 2019; Calosi & Mariani 2020; Lombardi & Ardenghi 2022; Buonocore 2022; Dorato & Morganti 2022. In contrast, Brown 2009; Wood 2010; Ruyant 2018; Dorato 2020; Pienaar 2021[a]; Adlam 2022 do acknowledge the iteration of relativity to some extent or other, although only Muciño, Okon & Sudarsky (2022, 13) identify the regress it gives rise to beyond the second order.

Second, one is tempted to complain that UIP cannot ultimately accomplish the feats it was designed for. For discussion, grant that we can relativise to one and the same system any number of times. Fine, one may then say, Wigner can find out which property S has relative to Friend, as these things stand relative to Wigner himself. And maybe he can find out which property S has relative to Friend, relative to Friend, as these things stand relative to himself; and so on for any level. But in order to *really* bridge the divide between their perspectives, Wigner would have to have the means to discover which property S has as relativised to Friend on *every* level. And, the interlocutor may continue, nothing in classical RQM ensures that the property of S when relativised to Friend on every level is the same as the property of S when relativised to Wigner on every level; hence, the worry about cross-perspective incoherences persists, too. But note that this objection itself rests on the idea that an absolute description can be recovered in the limit of the iterating process; a description (of the state of S relativised to Friend on every level 'at once') that is not itself relative. This is an assumption which classical RQM urges us to let go of.

Third, it should be understood that even with UIP on board, classical RQM is not a form of global relativism. The relativity it postulates is limited to state-dependent physical properties represented by the values of state-dependent variables (Laudisa & Rovelli 2019, 1.1). In contrast, the general framework of the pertinent quantum theory, including the algebra of observables, and of course the principles of RQM itself are supposed to enjoy absolute validity (cf. Van Fraassen 2010, 391; Dorato 2016, 253).

## 2.3 Quantum Relativism without the Iteration Principle

Interpretive considerations aside, there is also the systematic question of whether UIP *should* (or even has to) be embraced by advocates of RQM, or by quantum relativists more broadly speaking. This is a question I cannot pursue in any generality here. Nonetheless, let me take this section to pause and develop some considerations that bear on the issue.

The iteration principle is up and running as soon we take relatively possessed properties to be directly represented by variables of composites whilst holding fast to the relativity of the values of state-dependent variables. But the alert reader may already have noticed some wiggle room in the interpretation of the observables on composites. What Wigner can measure, it could be objected, are really the physical states of S and Friend (relative to him) at some later time $t_2$. But the (possible) fact that at $t_2$, S is in the 'up'-state and Friend's pointer indicates 'up' is not *identical* to the (possible) fact that S assumed the *up*-property relative to Friend in their earlier interaction. This opens up the logical space to maintain that only the former, but not the latter, is a relative issue. Thus, one could hold that the values of state-dependent variables (and the properties they represent) are always relative and yet there are *absolute* facts about the properties one system possesses relative to another. The immediate price to pay is that these latter facts cannot, on pain of contradiction, be directly represented by (the values of) state-dependent variables (of composites). In this sense, then, the quantum-mechanical description would come out incomplete. Moreover, insofar as the existence of relative properties between two systems is a matter of the dynamical state of their composite, this move would also amount to a restriction of Rovelli's 'main hypothesis' (according to which *no* absolute statements concerning dynamical states are possible).

However, we could still assume that inferential *connections* hold between absolute second-order facts and the values of state-dependent variables relative to particular references. This is important. For, recall that UIP is a supporting column rather than a freely spinning wheel: when imposing a restriction, one would do good not to re-import worries pertaining to the epistemic accessibility or consistency of events across references.[11] And the first way that comes to mind to ensure this is brute force: make it an additional postulate that the value which Friend's pointer variable takes on relative to Wigner if probed by him reliably tracks the (supposedly absolute)

---

[11] One natural idea to meet this challenge would be that relativisations come to an end when no external system that does not yet appear in the sequence of *relata* is left over. I think that this view faces problems, but I do not pursue the matter further here.

fact about the property S assumed relative to Friend in their earlier interaction.

This, in effect, is the strategy recently brought up for discussion by Adlam & Rovelli (2023), who entertain mainly two updates to classical RQM. First, they recommend to regard "the point-like quantum events or 'flashes' as absolute, observer-independent facts about reality" (ibid., 11). I take this to mean (at least) that the occurrence or not of an interaction between two systems itself becomes an absolute question. And second, they explicitly renounce the earlier assumption that comparisons between the interaction outcomes witnessed by two systems are themselves necessarily merely relative to references (ibid., 7). As a replacement, they introduce a new postulate – 'Cross-Perspective Links' (CPL) – designed to enable agreement between observers in an absolute sense[12]:

> "In a scenario where some observer Alice measures a variable *V* of a system *S*, then provided that Alice does not undergo any interactions that destroy the information about *V* stored in Alice's physical variables, if Bob subsequently measures the physical variable representing Alice's information about the variable *V*, then Bob's measurement result will match Alice's measurement result." (ibid.)

The possibility of an absolute comparison, in turn, seems to presuppose observer-independent facts about what Bob's and Alice's measurement results were in the first place. Nonetheless, the modified variant of RQM retains a mildly relationalist flavour insofar as quantum states are still regarded as reference-dependent; for, they are understood to encode the *joint* interaction history of the target system and some particular observer (ibid., 13).

All this is in keeping with Adlam's view that science requires absolute comparisons (2022, 9f.; 2024, 31) and that there are no good reasons to relativise facts about the outcomes of measurements. According to her, consideration of Wigner's Friend and Bell-Wigner scenarios should indeed motivate us to relativise quantum states, understood as mere "mathematical object[s] encoding predictions for the outcomes of measurements performed on a system" (2024, 13), but *not* the ontic state of a system, its "intrinsic description [...] at a given time" (ibid., 13; cf. ibid., 38). Importantly for our purposes, she also invokes the new version of RQM with CPL as one example for an interpretation that realizes this vision of hers.

---

12  Dieks (2022, 12f.) makes a similar suggestion, in turn tracing it back to Healey (2022).

One consequence of this is that RQM with CPL can no longer be motivated with recourse to no-go theorems for observer-independent facts about outcomes (which are quite commonly considered the strongest arguments in favour of a broadly 'relativist' interpretation of quantum theory; cf. Pienaar 2021, 2; Cavalcanti 2021, 2-4; Schmid, Ying, & Leifer 2023, 3).[13] Instead, Adlam (2023, 30) takes the updated version to *circumvent* such theorems by invalidating one of their premises. So, there are several respects in which the update comes out less relativist than classical RQM. In Adlam's words, RQM with CPL turns its back on any kind of "metaphysically radical non-absoluteness" (ibid., 1). To this extent, Stacey has a point when he speaks of a "de-relationalizing of Relational Quantum Mechanics" (2022, 2).

Let me summarise the little excursion undertaken in this section. There is logical room for a view on which state-dependent variables represent relative properties yet UIP does not hold. One way to set up such a view has been investigated into by Adlam & Rovelli. Perhaps unsurprisingly, the account they end up with comes out a good deal less relativist than classical RQM. A little more surprisingly, perhaps, we also find that one of the most popular strategies to motivate quantum relativism (on the basis of the Bell-Wigner theorems) no longer applies to it.

Whether or not RQM with CPL qualifies as a member of the 'quantum relativist' family is a matter of terminology. For present purposes, let it be granted that quantum relativists are not necessarily committed to UIP. The principle is of interest nonetheless, both insofar as it forms part of (the best reconstruction of) classical RQM and for independent systematic reasons. For, it is certainly a *possible* route to take for the relativist – in particular the one who is attracted to a strong reading of the idea that "there are no facts of the world per se, but only relative to observers" (Brukner 2015, 22).

---

13   Compare this to earlier claims that "these no-go theorems can be taken as direct evidence in favour of RQM" (Rovelli 2022, 1065).

## 3 (How) Do we Understand Rovelli's World?

### 3.1 Relations and Perspectives

Once a subject matter has been uncovered as relative, a natural instinct urges us to think about it in relational terms. Instead of saying that the car moves at 50 km/h, in our more careful moments we would be inclined to say that it moves at this speed *relative to us*. Typically, a statement of the former type would then be considered either confused or, more charitably, elliptical. In reality, it may be said, there *is no such thing* as simply having a certain velocity. This instinct is *so* natural, in fact, that Wright has called it the "basic indicator of folk relativism" (2008, 158). And what that means is that the ordinary way of being a relativist is to be a *relationalist* – someone who believes that the properties under scrutiny turn out to have higher adicity than expected heretofore (cf. Wright 2008, 158f.; Spencer 2016, 432f.). The relationalist fiddles with the *contents* of facts, because she believes these contents to involve reference to some additional parameter(s). Importantly, however, she holds that a genuine (potential) fact is one in which all *relata* have been made explicit, and always does obtain or not in an absolute manner. For the relationalist, it would at best be misleading loose talk to say that 'relative to me, it's a fact that the car moves at 50 km/h', since the only genuine fact in the vicinity is the (absolute) fact that the car moves at 50 km/h relative to me. For the relationalist, there is in reality no such thing as a fact that (only) obtains from a certain vantage point.

Sometimes, relationalism has been entertained as an explication of what relativism even is (Beardsley 1983, 265; Streiffer 2003, 4; cf. also Boghossian 2006$^b$, 13), and for quite some time, a majority of philosophers "roundly rejected" (Spencer 2016, 433) the notion that there was any coherent alternative. Arguably, physicists also isually think about relativity along relationalist lines.

Lately, though, the word has spread that relationalism is not the only game in town. The relativist could also adopt realism about perspectival facts (or 'perspectivism' for short).[14] Perspectivists leave the adicity of

---

14 Realism about perspectival facts (a metaphysical view) is not to be confused with the 'perspectival realism' that has attracted great interest in the philosophy of science (cf. Giere 2006; Massimi 2022). The latter revolves around the idea that *scientific knowledge* is perspectival in a substantial sense. Indeed, at least some proponents of this epistemological sort of perspectival realism are explicit that they take the *facts* themselves to be absolute properties and the contents of facts alone. Instead, they locate extra layers of relativity in the way in which these facts obtain: They think that some facts only obtain (or not) *from certain perspectives*, rather than absolutely. Put differently, they believe that whether or not some thing has a certain property sometimes is a perspective-dependent affair. On a perspectivist reading of Special Relativity (cf. Fine 2005; Lipman 2020), for instance, it may well be that an object has a certain velocity (*simpliciter*), and the claim that it does can be *literally* true, rather than confused or elliptical. It is just that whether or not that is the case is relative to a perspective (in this case, a reference frame).

The two forms of relativism give rise to two competing conceptions of the relativity of properties in classical RQM. Which of them was intended has always remained somewhat unclear, as will be documented in the course of the next sections. This should not surprise us. The distinction between relationalism and perspectivism is not particularly well-known, and may appear academic at first. Nonetheless, it matters.

## 3.2   How Relational is Relational Quantum Mechanics?

Although perspectival facts have recently gained currency, they are still fairly heterodox.[15] It is hence unsurprising that the relationalist temptation remains strong when we are exposed to RQM, and Rovelli himself was first to yield to this temptation. It was him, after all, who called his interpretation 'relational', and who declared that "[p]hysics is concerned with relations between physical systems" (1996, 1655). Explicating the ontology of RQM, he says that "[a]ll contingent (variable) properties are relational, namely they are *relations*: they refer to two systems, not one" (2022$^b$, 161; cf. also Rovelli 1996, 1649; Laudisa & Rovelli 2019, 3.1; Rovelli 2022$^a$, 1066). And entirely in line with the relationalist paradigm, he concluded that "there is no fact of the matter" (Laudisa & Rovelli 2019, 3.1; cf. also Rovelli 2005, 115) about the monadic properties of an object, so that ordinary claims about state-dependent physical properties are either to be reinterpreted "as elliptic

---

    (cf. Massimi 2018, 342), thus denying the metaphysical thesis of perspectivism. Thanks to an anonymous referee for raising this point.

15  The modal case might be an exception: it seems commonsensical to hold that different facts obtain relative to different possible worlds. Whether this counts as an instance of perspectivism in an interesting sense is debatable.

expressions for relational assertions" (Rovelli 1996, 1650) or otherwise "meaningless" (ibid., 1666).

None of this is to deny that *other* statements of his point in a very different direction (I come to that in a minute). But it explains why many have taken RQM to be a relationalist theory, and more specifically one which prescribes a replacement of monadic properties by *dyadic* relations (cf. Candiotto 2017; Laudisa 2019, 216; Glick 2021, 53; Martin-Dussaud 2021, 8; Calosi & Mariani 2021, 13; Vidotto 2022, 3f.; Dorato & Morganti 2022, 2; Buonocore 2022, 5; Lombardi & Ardenghi 2022, 11). The conclusion that the world so described is a sort of relational structure is but a small step away:

> "The physical world must be described as a net of interacting components [...] The state of a physical system is the net of the relations it entertains with the surrounding systems. The physical structure of the world is identified as this net of relationships." (Laudisa & Rovelli 2019, 3.1)

It is this vision that is is at work when RQM is read as a variety of Ontic Structural Realism (Candiotto 2017), or when formal 'fact-nets' of dyadic relations are proposed as the proper language for Rovelli's interpretation (Martin-Dussaud et. al., 2023).

But there is something off about all this. Given UIP, a description of a physical situation in terms of binary predicates is just as reference-dependent and incomplete as a description in terms of unary predicates. In this respect, a binary relational fact-net – whatever its pragmatic merits – can be no better a representation of the world of classical RQM than one in terms of the old unary predicates. This representational predicament only mirrors a metaphysical one: As believers in classical RQM, we cannot simply think of the world as a structure of dyadic relations, because which of them do obtain or not would itself be a relative affair. And given the relationalist's irrealism about perspectival facts, this means that there is no more a fact of the matter about these relations than there is about *the* velocity of an object. If talk in terms of unary predicates is, strictly speaking, meaningless (because there are no corresponding monadic properties), then talk in terms of binary predicates is, too (because there are no corresponding dyadic properties). Precisely the same issue afflicts triadic, tetradic, or pentadic relations. There can be no "net of relationships" that describes *the* structure of the world, because whether or not the relationships it describes do indeed obtain is always itself going to be a reference-dependent question.[16] Whenever you hand me such a net, my inner Rovelli points to the iteration principle and proclaims: "Well! This is at most what the world is like relative to some system!"

A moment of reflection reveals that the combination of relationalism with UIP is a recipe for disaster. The iteration principle has it that any attribution of a state-dependent property to some system relative to any sequence of *relata* ought to be regarded as, yet again, relative; or, turning things around, that no purported fact about such matters obtains absolutely. But relationalists do not entertain facts that obtain relative to vantage points, either. The implication is that no facts about the dynamical state of anything obtain *at all* (neither absolutely nor relatively).[17] This is hardly a convenient basis to build an account of physical reality on. Another way of putting the same point is that for any state-dependent property $\varphi$ of whichever arity, the relationalist and iterator of relativity is forced to say that there really is 'no such thing' as $\varphi$-ing. She is left without state-dependent properties to play with.[18]

I wish to make it very clear that I am not here begging the question against the quantum relativist by dogmatically requiring that there be an absolute level of description.[19] It is the combination of UIP with *relationalism* that creates trouble. Relationalism implies that anything that is the case is the case absolutely, which yields absurd results if no absolute level exists.

As is, classical RQM cannot be given a sensible relationalist reading. This has two immediate implications. First, it puts further pressure on the idea that the relativity of dynamical properties in RQM is continuous with, or conceptually coheres particularly well with, 'relationalism' as a general key feature of modern physics (Vidotto 2022; Martin-Dussaud 2021). Second, we face a dilemma: We either have to give up on relationalism or hedge in UIP.

---

16  Recall the idea expressed in *Helgoland* that even relational structures themselves merely exist "in relation to something else" (2021$^b$, 127).

17  One may think that those who deny that the metaphysical 'catalogue of what exists' contains 'facts' (e. g., Betti 2015) could simply accept this conclusion. But this would be a misunderstanding, because facts talk is dispensable in the above argument. The same problem could be phrased in terms of what is the case (either absolutely or relatively), the conclusion being that nothing (to do with the dynamical states of physical systems) is the case at all.

18  Kölbel (2011, 24f.) makes an analogous point about a view on which non-relativised statements in a certain domain fail to express complete propositions.

19  Thanks to an anonymous referee for pushing me on this issue.

Again, a restriction of the principle should respect its original motivations, and avoid worries of cross-perspective consistency.

Towards the end of a lesser-known paper, Rovelli says the following:

> "The precise time at which the measurement happens can be observed, and can be operationally defined [...], but not by O. Rather, by the third system (O') making measurements on the S-O system. Simultaneous use of these different levels does not lead to any contradiction. And there is no regression at infinity, unless we inquire about absolute reality, an inquiry which, I suspect, is illegitimate." (1998, 1042)

Relationalism is the version of relativism that does not – cannot – abstain from inquiring into absolute reality (cf. Baghramian 2004, 43). Whether the regress Rovelli has in mind here is exactly the one I have discussed, I leave to the reader to judge.

### 3.3 Towards Perspectival Quantum Mechanics

One natural conclusion to draw from the foregoing is that classical RQM requires, or at least favours, perspectivism. All the apparently relationalist talk notwithstanding, there is ample textual support for this move. Recall, for instance, Rovelli's suspicion that inquiries into absolute reality are 'illegitimate', and that there is no "view from nowhere" to be had (Di Biagio & Rovelli 2022, 11). In the recent past, 'relations' were more and more superseded by 'relative facts', wherefore Pienaar proposed "to rename RQM the 'Relative-facts interpretation of quantum mechanics'" (2021[a], 14). At one point, Di Biagio & Rovelli explicitly say that a "previously 'obvious' notion – absolute facts – may in fact be inappropriate to describe the world" (2022, 7). All this is very much in line with the popularity of the concept of 'relative facts' in quantum relativism (cf. Brukner 2017, 113f.; Pienaar 2021[b], 1; Healey 2022, 9-12).

There is a good case to be made that perspectivism is the natural home for classical RQM, and for the quantum relativists more generally. What I would like to bring out in the remainder is that this commitment is not to be taken too casually, and, more importantly, that it only marks the beginning of the conceptual work that needs doing.

The first thing to register is that the tenability of quantum relativism, if indeed it is to be cashed out in perspectivist terms, is wedded to the fate of a framework whose coherence remains disputed (for particularly fundamental opposition, cf. Moore 1997, 50). But even friends of perspectival facts concede that they come with theoretical costs. A landmark analysis due to Fine (2005) has convinced many that perspectivists are forced to abandon at least one of the following cherished principles:

| | |
|---|---|
| NEUTRALITY | There is no metaphysically privileged perspective. |
| COHERENCE | Reality is not constituted by facts with contradictory content. |
| ABSOLUTISM | The constitution of reality is an absolute matter. It is not itself relative to a perspective. |

Since the notion of a unique, coherent reality that does not privilege any particular vantage point is appealing, this predicament has been referred to as "Fine's trilemma" (Loss 2018). In a nutshell, it arises simply because facts that supposedly obtain relative to different standpoints – facts the perspectivist takes seriously – may well be inconsistent with one another.[20]

Several types of perspectivism can be distinguished as a function of the principle they choose to discard. According to *privileged-perspective realism*, we should give up NEUTRALITY and accept that reality is made up by the facts that obtain relative to one particular preferred standpoint. According to *fragmentalism*, COHERENCE is false, as reality fans out into various internally (but not mutually) coherent collections of obtaining facts. Finally, *external relativism* has it that contrary to ABSOLUTISM, the very constitution of reality is a perspective-dependent matter.

Neither Rovelli nor those who have endeavoured to flesh out his conception of 'relative facts' (Ruyant 2018, Healey 2022) have engaged with Fine's trilemma, or otherwise clarified which type of perspectivism (if any) they have in mind. When I point this out, people sometimes reply that it is *evident* which version best suits classical RQM. Unfortunately, they disagree over which one it is. Even granting that privileged-perspective realism sorts ill with Rovelli's egalitarian stance (cf. Rovelli 2022[a], 1067), two options remain on the table. Some seem to consider fragmentalism unfit for RQM because RQM is all about *dissolving* apparent incoherences. Others think that external relativism is outright unintelligible, or that it would be against the spirit of RQM to relegate the facts that obtain relative to other systems to a realm 'outside of' (one's own) reality. The disagreement is also manifest in the literature. Dieks (2022, 17) generally recommends fragmentalism as a promising avenue for quantum relativism. In remarkable contrast, Healey puts forward the (somewhat perplexing) claim that *unlike* RQM, fragmentalism "does not involve relative facts" (2022, 10).

---

20  As Fine acknowledges, arguments of similar nature go back at least to McTaggart (1908).

It is not obvious which type of perspectivism is the right one to spell out classical RQM. Perhaps one may be tempted to argue that the question is *irrelevant* for the project of interpreting quantum theory, that it belongs to the cleaning-up operations that can safely be left to metaphysicians to do later. But again, I am not convinced that this would be apposite, for at least two reasons.

Firstly, if quantum relativists make use of the non-standard notion of 'relative facts', the burden is on them to clarify it, and to reveal how sense can be made of it. After all, arguments like Fine's seem to show that the notion gives rise to inconsistencies when put to work naïvely. Or, to approach the same point from the opposite angle: if they do not want to be relationalists, quantum relativists cannot just piggyback off relationalist intuitions and take for granted that relativisation actually removes the apparent contradictions it is invoked to remove. They should explain how this is supposed to work, given their realism about the incompatible facts at issue, rather than trusting others to make sense of it. But this essentially just means facing Fine's trilemma.

Secondly, the conceptual elaboration of quantum relativism also serves to bring out further commitments in the interest of an accurate cost-benefit analysis. Each of the different types of perspectivism face their own set of drawbacks, or even charges of incoherence (cf. Deng 2017, 1119). And to safeguard their positions, perspectivists have employed an arsenal of rather fanciful metaphysical notions. Fine claims that both fragmentalism and external relativism are, in different ways, committed "to a distinction between a single comprehensive über-reality and a plurality of more particular realities" (2005, 282), conceding that "it is very hard to say what this distinction comes to" (ibid.). Fragmentalists have relied on distinctions between how things are and how they are *in reality* (Fine 2005, 262), between facts that obtain and those that *constitute reality* (Iaquinto 2019, 696), or between facts that *exist* and those that *obtain* (Iaquinto 2020, 577). Merlo (2022) can be taken to argue that some such differentiation is indispensable for fragmentalism to be tenable. But it is not clear whether the quantum relativists would be happy to make these notions their own.

Apart from the species of perspectivism, the fate of UIP is another issue that deserves attention. I have throughout defended the claim that it forms part of classical RQM in its most adequate reconstruction. But this does not seem to be common ground among those who offer perspectivist takes on the interpretation. For instance, Healey's (2022, 13-15) discussion of cross-perspective communication in the Wigner's friend case seems to rely on the presupposition that there are absolute facts about the properties one system assumes relative to another. Even if Friend actually received outcome *up*, Healey says, Wigner's quantum state still encodes a non-zero probability for the possibility that he will find Friend in the state $|F_{down}\rangle$. Hence, Healey claims, Wigner's "relative outcome is completely uncorrelated to F[riend]'s relative outcome of her measurement" (ibid., 15). But this is a refusal to engage with Rovelli's story. On that story (Rovelli 1996, 1651f.), there *is* no absolute fact about what is the case from Friend's perspective, and Wigner's quantum state correctly outputs the probabilities for Friend's measurement outcome *as is relative to Wigner*. Smerlak & Rovelli (2007, 440f.) repeat exactly this point in anticipation of exactly the same criticism. This leads me to conjecture that Healey, at least implicitly, presupposes that relativity does not iterate.

The case of Ruyant (2018) is less clear. He proposes modelling RQM by means of a "relativist conception of truth, where truth is a two place predicate indexed to an observer" and even "second-order statements, such as 'p is true-for-O' [...] only [have] a truth-value relative to an observer" (2018, 447). Evidently, Ruyant acknowledges the second layer of relativity. However, his way of setting up the issue at least suggests that he does either not embrace or not consider *higher* orders. Why else the emphasis on *second*-order statements? Why a *two*- rather than three- or four-place predicate?[21]

Whether my exegeses are correct is ultimately neither here nor there. What matters is that quantum relativists should clarify their stance towards

---

21  In fact, it would be natural for perspectivists to stick to the monadic truth predicate. They can maintain that a (non-relativised) statement can be simply *true*, although whether or not it is depends on a standpoint. Yet again, the matter hinges on the details of the proposal.

UIP.[22] If they do not embrace it, relationalism is back in the game. If they do, they commit to an infinite iteration of perspective-dependencies. Again, this may be expected to have an influence on the overall cost-benefit balance. For one thing, it calls into question the claim that RQM is particularly parsimonious (Laudisa & Rovelli 2019). In a way, the resultant metaphysical structure is breathtakingly complex. We started out with a world described by the facts, period; now, we are told that what is the case is a matter of perspective, and indeed that what is the case relative to a perspective is itself a matter of perspective, and so on *ad infinitum*.

Some may even think that the iteration principle renders the interpretation indefensible. That philosophers take exception with regresses of relativity is a recurring theme in the literature on relativism. Putnam remarked that they make "our grasp on what the position even means [begin] to wobble" (1981, 121). Both Burnyeat (1976, 192f.) and Boghossian (2006[a], 56) worry that the iterating relativist is in some way committed to the attempt of entertaining propositions much too complex for the human mind. I suspect that Dorato expresses a common sentiment when he says that "the coherence of any form of relationism presupposes the existence of perspective-independent facts about what the facts are from each perspective" (2016, 252). It is true that the arguments developed in favour of this stance, if any are given, are often sketchy or uncharitable (cf. Bennigson (1999) or MacFarlane (2014) for replies). But how this debate plays out in the specific context of quantum relativism is an unaddressed question; and it at least deserves attention that the fate of classical RQM, or any quantum relativist view, should depend on it.[23]

---

22  One may think that whoever claims that there is no 'view from nowhere' is committed to the iteration principle. But this is not so clear, because the metaphor isn't. As soon as a single fact *f* obtains from some perspective but not from another, there is a sense in which one cannot give a complete description of the facts without choosing a standpoint if one requires such a description to specify whether or not *f* obtains (and not only whether it obtains relative to this or that perspective). I owe this observation to a remark by Martin Lipman.

23  Here is one final can of worms: Granting that perspectivist RQM (with UIP) is a coherent description of what the world is possibly like, we still need to understand how its metaphysics connects with our life as embedded agents, with our assertions, beliefs, experiences, and expectations. If my states of consciousness are intimately connected to the dynamical state of my body, and if my body only ever has dynamical states relative to an infinite sequence of *relata*, it plausibly follows that my experiences are relative in the same way. If so, there is no absolute fact about which state of consciousness I am going through at one particular instant; instead, the matter depends on a never-ending series of vantage points. Take my experience of typing words on a computer, feeling a little confused. Relative to whom should I take myself to be having this experience? If mental states

## 4    Conclusions

My aim in this paper was to draw attention to the issue of the iteration of relativity in RQM, and in relativist approaches to quantum theory more generally. I argued that the best reconstruction of 'classical' RQM incorporates the Unrestricted Iteration Principle, and I described the work the latter is supposed to do to dispel concerns about the accessibility and compatibility of events across references. Subsequently, though, it was shown that UIP is incompatible with a relationalist reading of RQM, a reading that is both common and conservative.

The finding confirms the suspicion that Rovelli's interpretation rather relies on the more extravagant framework of perspectivism. But while it is easily said that reality is perspectival, filling in the conceptual details and developing the catchy slogan into a coherent account is hard work. Among the open questions that I have argued need addressing are the type of perspectivism at issue as well as the fate of the iteration principle. The general point that emerges from this is that the theoretical costs of a fully fleshed out perspectivist version of RQM may be higher than many who were attracted by the catchphrase would have hoped. If so, perhaps they turn to Adlam's & Rovelli's recent proposal, and rethink RQM as a less relativist, event-based interpretation.

---

supervene on the state of my body, there is presumably no qualitative difference between the experiences I have relative to myself and those I have relative to my neighbour. Is the best thing I can say that I am having my experience relative to *some* sequence of *relata*? Which facts do I gain access to by means of my perceptions then? Is all I can say that I can see that *p* is the case relative to *some* sequence of *relata*? Is that good enough to do science?